\documentstyle[prl,aps,multicol,epsfig]{revtex}

\begin{document}

\title{A Conformal Hyperbolic Formulation of the Einstein Equations}
\author{Miguel Alcubierre${}^{(1)}$, Bernd Br\"ugmann${}^{(1)}$, Mark
Miller${}^{(2)}$ and Wai-Mo Suen${}^{(2,3)}$}

\address{${}^{(1)}$Max-Planck-Institut f\"ur Gravitationsphysik \\
Albert-Einstein-Institut \\
Schlaatzweg 1, D-14473, Potsdam, Germany}

\address{${}^{(2)}$McDonnell Center for the Space Sciences \\
Department of Physics,
Washington University, St. Louis, Missouri 63130}

\address{${}^{(3)}$Physics Department \\
Chinese University of Hong Kong,
Hong Kong}

\date{\today}
\maketitle

\begin{abstract}
We propose a re-formulation of the Einstein evolution equations that
cleanly separates the conformal degrees of freedom and the
non-conformal degrees of freedom with the latter satisfying a first
order strongly hyperbolic system.  The conformal degrees of freedom
are taken to be determined by the choice of slicing and the initial
data, and are regarded as given functions (along with the lapse and
the shift) in the hyperbolic part of the evolution.

We find that there is a two parameter family of hyperbolic
systems for the non-conformal degrees of freedom for a given set of
trace free variables.  The two parameters are uniquely fixed if we
require the system to be ``consistently trace-free'', i.e., the time
derivatives of the trace free variables remains trace-free to the
principal part, even in the presence of constraint violations due to
numerical truncation error.  We show that by forming linear
combinations of the trace free variables a conformal hyperbolic system
with only physical characteristic speeds can also be constructed.

\end{abstract}
\begin{multicols}{2}

\paragraph {INTRODUCTION.} 

With the advent of large amounts of observational data from high-energy
astronomy and gravitational wave astronomy, general relativistic
astrophysics --- astrophysics involving gravitational fields so strong
and dynamical that the full Einstein field equations are required for
its accurate description --- is emerging as an exciting research area.
This calls for an understanding of the Einstein theory in its
non-linear and dynamical regime, in order to study the physics of
general relativistic events in a realistic astrophysical environment.
This in turn calls for solving the full set of Einstein equations
numerically.  However, the complicated set of partial
differential equations present major difficulties in all of these
three tightly coupled areas: the understanding of its mathematical
structure, the derivation of its physical consequences, and its
numerical solution.  The difficulties have attracted a lot of recent
effort, including two ``Grand Challenge''
\cite{Alliance98,Nasa98} efforts on the numerical studies of black
holes and neutron stars, respectively.

One major obstacle in solving the Einstein equations numerically is
that we lack a complete understanding of the mathematical structure of
the Einstein equations.  The difficulties in numerically integrating the
Einstein equations in a stable fashion have motivated intense effort
in rewriting the Einstein equations into a form that is explicitly
well-posed
\cite{Friedrich81a,Friedrich81b,Choquet83,Friedrich85,Bona92,Bona94b,Choquet95,Fritelli95,Abrahams95a,Friedrich96,MVP95,Abrahams96a,Bona97a,Abrahams97b,Bona98b,Anderson99} (for an excellent overview see \cite{Reula98a}).
The main idea has been re-writing the six space-space components of
the Einstein equations into a first order hyperbolic system.  These
space-space parts of the Einstein equations are dynamical
evolution equations, while the time-time and space-time parts of the
Einstein equations are (elliptic) constraint equations.

The central question we raise in this communication is: In order to
enable an accurate and stable numerical integration of the full
set of the Einstein equations, what part of the system should be taken
to form a hyperbolic system?

Our question is motivated by two observations.  First, there are many
recent proposals on re-formulating the six space-space parts of the
Einstein equations into a first order hyperbolic system
\cite{Bona94b,Choquet95,Fritelli95,Abrahams95a,Friedrich96,MVP95,Abrahams96a,Bona97a,Abrahams97b,Bona98b,Anderson99}.
Three of the hyperbolic formulations have been coded up for numerical
treatment (to the best of our knowledge), namely, the York et. al.
formulation \cite{Choquet95,Abrahams96a,Abrahams97b} (see,
e.g., \cite{Scheel97,Scheel98a}), the Bona-Masso formulation
\cite{Bona97a} (see e.g., \cite{Bona98b}), and Friedrich's formulation
\cite{Friedrich81a,Friedrich81b,Huebner93,Huebner96,Huebner98,Frauendiener98a,Frauendiener98b}
(which is rather different from the first two formulations in its use
of a global conformal transformation of the four-metric to compactify
hyperboloidal slices).  However, in all these cases, the numerical
integration of the first order hyperbolic system consisting of the six
space-space components of the Einstein equations so far have not lead
to a substantial improvement over those using the traditional ADM
\cite{Arnowitt62} evolution equations. This is despite the original
hope that the well-posedness of the hyperbolic formulations leads to
an immediate numerical advantage.

The second observation is that there have been various attempts in
re-writing the traditional ADM form of the evolution equations by
separating out the conformal degree of freedom, beginning with
Nakamura et. al. \cite{Shibata95} (see references cited therein).
Lately this has received much attention with \cite{Baumgarte99}
reporting that a variant of the approach leads to highly stable
numerical evolutions.  A detailed study of the approach using
gravitational wave systems carried out by our group
\cite{Alcubierre99a} confirmed that the approach has advantages over
the standard ADM formulation.  We find that the approach yields
results with accuracy comparable to that obtained by the standard ADM
formulation with the K-driving technique \cite{Balakrishna96a} for
weak to medium waves, and has better stability properties especially
in the case of strong fields that needs high resolution with ADM
\cite{Alcubierre99b} (see also \cite{Arbona99}).

These two observations motivated us to study the possibility of a
formulation that separates out the conformal degree of freedom
in the 6 evolution equations, while requiring the remaining 5
equations governing the non-conformal degrees of freedom to form a
first order hyperbolic system.

A re-cap of the various components of the Einstein equations is in
order for a clearer discussion of our approach. In the standard ADM
3+1 formulation, the Einstein equations are broken into (a) the Hamiltonian
constraint equation (the time-time part),
\begin{equation}
\label{eq:ham}
H = {}^{(3)}R + K^2 - K_{ij} K^{ij} - 16 \pi {\rho}_{{}_{ADM}} = 0,
\end{equation}
(b) the 3 momentum constraint equations (the time-space part)
\begin{equation}
\label{eq:mom}
H ^ i = \nabla_j K^{ij} - \gamma^{ij} \nabla_j K - 8 \pi j^i = 0 ,
\end{equation}
where ${\rho}_{{}_{ADM}},j^i,S_{ij},S=g^{ij} S_{ij}$ 
are the components of the stress
energy tensor projected onto the 3-space, and (c)
the 6 evolution equations (the space-space part) given
as 12 first order equations
\begin{eqnarray}
\partial_{\hat{t}} g_{ij}  &=& - 2 \alpha K_{ij} ,
\label{eq:gij} \\
\partial_{\hat{t}} K_{ij} &=& -\nabla_i \nabla_j \alpha + \alpha (
R_{ij}+  
K\ K_{ij} -2 K_{im} K^m_j \nonumber \\ &\ & 
- 8 \pi ( S_{ij} - \frac{1}{2} g_{ij}S )
- 4 \pi {\rho}_{{}_{ADM}} g_{ij})  ,
\label{eq:kij}
\end{eqnarray}
where $\nabla_i$ denotes a covariant derivative with respect to the
3-metric $g_{ij}$, $\partial_{\hat{t}}$ stands for $\partial_t -
{\cal{L}}_{\beta}$ with $ {\cal{L}}_{\beta}$ 
being the Lie derivative with respect to
$\beta ^i$, and $R_{ij}$ is the Ricci curvature of the 3-metric.  In
the ADM formulation, Eqs. (\ref{eq:gij} ,\ref{eq:kij}) are
used to evolve the 12 variables {$K_{ij}$, $g_{ij}$} forward in time
for given lapse $\alpha$ and shift vector $\beta ^ i$.  The constraint
equations are automatically satisfied if \{$K_{ij}$, $g_{ij}$\}
satisfy them on the initial time slice.  However, in numerical
evolutions the constraints will be violated due to truncation error.
One major difficulty in numerical relativity is that the constraint
violations often drive the development of instabilities, at least in
the case of numerical evolution using the standard ADM equations
(\ref{eq:gij}, \ref{eq:kij}).

In the hyperbolic re-formulations of the evolution equations
\cite{Choquet95,Fritelli95,Abrahams95a,Friedrich96,Abrahams96a,Bona97a,Abrahams97b,Bona98b},
one makes use of the constraint equations
(\ref{eq:ham}),(\ref{eq:mom}), and introduces additional variables
(e.g., $d_{ijk} = g_{ij,k}$ or its linear combinations) to cast Eqs.
(\ref{eq:gij} , \ref{eq:kij}) into a first order strongly hyperbolic
system (often the symmetric hyperbolic subclass).  (More variables
would have to be introduced for formulations involving higher
derivatives \cite{Choquet95,Abrahams95a}.)  However, we note that
hyperbolicity is often shown only under the assumption that some of
the variables involved in the evolution equations, in particular the
lapse $\alpha$ and the shift $\beta ^j $, are considered as given
functions of space and time.  In actual numerical evolutions with no
pre-determined choice of spacetime coordinates, $\alpha$ and $\beta ^j
$ have to be given in terms of the variables \{$K_{ij}, g_{ij},
d_{ijk}$\} (e.g., $\alpha , \beta ^j $ determined in a set of elliptic
equations involving \{$K_{ij}$, $g_{ij}$, $d_{ijk}$\}).  In the
Bona-Masso formulation \cite{Bona97a}, the lapse can be part of the
hyperbolic system for some choice of slicings (while the inclusion of
the shift into the hyperbolic system severely restricts the class of
applicable shifts).  In \cite{Choquet95,Abrahams95a}, in addition to
the lapse and the shift, the trace of the extrinsic curvature, $K =
g^{ij} K_{ij}$, is also regarded as a given function ($K$ is used to
specify the slicing, e.g., $K =0$ for maximal slicing).  The point we
want to bring out here is that in all of the existing hyperbolic
re-formulations of the Einstein evolution equations, part of the
quantities \{$K_{ij}, g_{ij}, d_{ijk}, \alpha , \beta ^j $\} are
considered to be given, while others are evolved using hyperbolic
equations.

In the following, we present a formulation in which the non-conformal
degrees of freedom are separated out for hyperbolic evolution.

\paragraph {FORMULATION.} 
For the evolution of the three-geometry, the conformal degree of
freedom is represented by $g$ (the determinant of the spatial 3-metric
$g_{ij}$), its spatial derivative $g_{,i}$ and its time derivative $K$
($K=-1/(2g \alpha ) \partial_{\hat{t}} g$).  For the non-conformal
degrees of freedom, we define 
\begin{eqnarray}
\label{eq:g}
\tilde{g} _{ij} &=& g_{ij}/{g^{1/3}} ,
\\
\label{eq:k}
\tilde{A} _{ij} &=& (K _{ij} - \frac{1}{3} {g}_{ij} K)/{g^{1/3}},
\\
\label{eq:d}
\tilde{D}^{ij} {}_{k} &=& \tilde{g}^{ij} {}_{,k}  .
\end{eqnarray}
$\tilde{g} _{ij}$ has unit determinant, and $\tilde{A} _{ij}$ is the
rescaled trace-free part of $K _{ij}$.  All indices of tilde
quantities are raised and lowered with the conformal 3-metric
$\tilde{g}_{ij}$.  We note that $\tilde{D}^{ij} {}_{k}$ is trace-free
with respect to the indices $(i,j)$.  We take \{$\tilde{g}^{ij},
\tilde{D}^{ij} {}_{k}, \tilde{A} ^{ij}$\}, or their covariant
component counterparts, to represent the non-conformal degrees of
freedom.

In the following we develop a first order hyperbolic system for the
non-conformal degrees of freedom, under the simplifying assumption
that the 5 conformal degrees of freedom \{$g, g_{,i}, K$\} and the
gauge choice functions \{$\alpha, \beta^i$\} can be regarded as given
functions of space and time.  Note that these variables cannot be
specified independently of each other. A concrete example is that of
maximal slicing, $K = 0$, and vanishing shift, $\beta^i = 0$, in which
case both $g$ and $g_{,i}$ are part of the initial data (time
independent), and are therefore truely given functions in the
numerical evolution. In other cases, with $K$ given to specify the
slicing, it involves a non-trivial time integration to determine $g$
(from the definition of $K$ in terms of the time derivative of $g$).

We now discuss hyperbolicity of the evolution of the non-conformal
variables, \{$\tilde{g}_{ij}, \tilde{D}^{ij} {}_{k}, \tilde{A}
^{ij}$\}, by examining the principal part of the evolution equations,
which is the part that decides about strong hyperbolicity of the
system \cite{Gustafsson95}.  To obtain the principal part we drop all
terms that can be expressed by (1) the variables \{$\tilde{g}^{ij},
\tilde{D}^{ij} {}_{k}, \tilde{A} ^{ij} $\} themselves, and (2)
spacetime functions that are regarded as given, i.e. \{$\alpha, \beta
^i , g , g_{,i}, K $\} and their space and time derivatives.
We have
\begin{eqnarray}
\partial_{\hat{t}} \tilde{g}_{ij} &\approx& 0 ,
\label{eq:gtilde}
\\
\partial_{\hat{t}} \tilde{D}^{ij}{}_{k} &\approx& 2 \alpha 
\tilde{A}^{ij}{}_{,k},
\label{eq:ddot}
\\
\partial_{\hat{t}} \tilde{A}^{ij} &\approx& \alpha g^{1/3} (R ^{ij}
- \frac{1}{3} g^{ij} R) , 
\label{eq:ktidle}
\end{eqnarray}
where $\approx$ represents ``equal up to principal part'', and where
for the evolution equation of $\tilde{D}^{ij}{}_{k}$ we have used that
spatial derivatives $\partial_i$ and the time derivative
$\partial_{\hat{t}}$ commute.

To evaluate $R^{ij}$ and $R$ in (\ref{eq:ktidle}), we use
\begin{eqnarray}
\label{eq:rij}
R^{ij} &\approx& g^{-2/3} \tilde{R}^{ij}  \\ &\approx&
\frac {1}{2} g^{-2/3} (\tilde{g}^{kl} {\tilde{D}^{ij}}{}_{k,l} -
\tilde{g}^{il} {\tilde{D}^{jk}}{}_{l,k} -
\tilde{g}^{jl} {\tilde{D}^{ik}}{}_{l,k}) ,
\end{eqnarray}
where the relation
\begin{equation}
\label{eq:gd}
g_{kl} {g^{kl}}_{,i} = -g_{,i}/g \approx 0 ,
\end{equation}
and the spatial derivatives of it have been used.  We obtain
\begin{eqnarray}
\label{eq:kdot}
\partial_{\hat{t}} \tilde{A}^{ij} &\approx& \frac{1}{2} \alpha g^{-1/3}
 ( \tilde{g}^{kl} \tilde{D}^{ij}{}_{k,l} - \tilde{g}^{il}
 \tilde{D}^{jk}{}_{l,k} - \tilde{g}^{jl} \tilde{D}^{ik}{}_{l,k} \nonumber \\
&~& + \frac{2}{3} \tilde{g}^{ij} \tilde{D}^{kl}{}_{k,l}).
\end{eqnarray}

To make the non-conformal system strongly
hyperbolic, one can add a combination of the momentum constraint
to (\ref{eq:ddot}).  To principal part the momentum
constraint (\ref{eq:mom}) is
$H ^ i \approx g^{-1/3} \tilde{A}^{ij}{}_{,j} $.
We obtain
\begin{eqnarray}
\label{eq:ddot2}
\partial_{\hat{t}} \tilde{D}^{ij} {}_{k} &\approx& 2 \alpha 
\tilde{A}^{ij}{}_{,k} - 2 \alpha g^{1/3} (\tilde{g}^i_k H^j + 
                                          \tilde{g}^j_k H^i)\\
\label{eq:ddot3}
{}  &\approx& 2 \alpha (\tilde{A}^{ij}{}_{,k} -
\tilde{g}^i_k \tilde{A}^{jl}{}_{,l} - \tilde{g}^j_k \tilde{A}^{il}{}_{,l}) .
\end{eqnarray}
An energy norm can be constructed for the system: 
\begin{equation}
\label{eq:e1}
E = \int \tilde{g}^{ij} \tilde{g}_{ij}+ \tilde{A}^{ij}
\tilde{A}_{ij} + \frac{1}{4} g^{-1/3} \tilde{D}^{ij}{} _{k} 
\tilde{D}_{ij}{} ^{k}.
\end{equation}
It is straightforward to demonstrate using (\ref{eq:gtilde}),
(\ref{eq:kdot}), and (\ref{eq:ddot3}) that $\partial_t E$ is a total
derivative up to terms that can be expressed by the variables
\{$\tilde{g}^{ij}, \tilde{A}^{ij}, \tilde{D}^{ij}{}_{k}$\} themselves.
One can also show directly that the characteristic
metric of the system (\ref{eq:gtilde}), (\ref{eq:kdot}), and
(\ref{eq:ddot3}) has a complete set of eigenvectors with real eigen
values.  The system is similar to but {\it not} contained in the one
parameter family of the hyperbolic systems constructed in
\cite{Fritelli95}.

Next we go one step beyond hyperbolicity.
We make the following observations:

\noindent
(i) Since $\tilde{A}^{ij}$ and $\tilde{D}^{ij}{}_{k}$ are trace-free,
one can add a term $\epsilon_1 \alpha g^{-1/3} \tilde{g}^{ij} H$ 
to (\ref{eq:kdot}), and a term
$\epsilon_2 \alpha \tilde{g}^{ij} H_k$ to (\ref{eq:ddot3}) 
without affecting the
hyperbolicity.  We have therefore a two parameter family of hyperbolic
evolution equations (without making a variable change).

\noindent
(ii) With these two terms added respectively to (\ref{eq:kdot}) and
(\ref{eq:ddot3}), the trace of the principle parts of the RHS's of the
equations are $3 \epsilon_1 \alpha \tilde{D}^{ks}{}_{k,s}$
(proportional to the principal part of the Hamiltonian constraint),
and $\alpha (3 \epsilon_2 - 4) \tilde{A}_k{}^l{}_{,l}$ (proportional
to the principal part of the momentum constraint), respectively.  On
the other hand, the LHS of the equations, $\partial_{\hat{t}}
\tilde{A}^{ij}$ and $\partial_{\hat{t}} \tilde{D}^{ij} {}_{k}$ are
trace-free to the principal order.  This means that truncation error
in the numerical evolution which leads to a violation of the
constraints will drive $\tilde{A}^{ij}$ and $\tilde{D}^{ij}{}_k$ to
evolve away from being trace-free, even up to the principal order.

\noindent
(iii) We therefore propose to fix the freedom in the parameters $\epsilon_1$
and $\epsilon_2$ by requiring the system to be ``consistently
trace-free'', i.e., $\epsilon _1 = 0$ and $\epsilon _2 = 4/3$,  so
that the equations are trace-free to principal order consistently.
Hence (\ref{eq:kdot}) for $\tilde{A}^{ij}$ is left unchanged, but
\begin{eqnarray}
&& \partial_{\hat{t}} \tilde{D}^{ij} {}_{k} 
\nonumber
\\
&\approx& 2 \alpha
\tilde{A}^{ij}{}_{,k} - 2 \alpha g^{1/3} (\tilde{g}^i_k H^j +
                                          \tilde{g}^j_k H^i) 
+ \frac{4}{3} \alpha \tilde{g}^{ij} H_k 
\label{eq:ddot4}
\\
\label{eq:ddot5}
&\approx& 2 \alpha (\tilde{A}^{ij}{}_{,k} - \tilde{g}^i_k
\tilde{A}^{jl}{}_{,l} - \tilde{g}^j_k \tilde{A}^{il}{}_{,l} + 
\frac{2}{3} \tilde{g}^{ij} \tilde{g}_{km}  \tilde{A}^{ml}{}_{,l} ) .
\end{eqnarray}
The system \{(\ref{eq:gtilde}),(\ref{eq:kdot}),(\ref{eq:ddot5})\}
forms a strongly hyperbolic system with the same energy norm
(\ref{eq:e1}). 

\noindent
(iv)The remaining freedom in constructing conformal-hyperbolic systems
that are ``consistently trace-free'' is through forming linear
combinations of the variables.  There are clearly infinite choices.
Here we show for example a linear combination that leads to a system
with only physical characteristic speeds, a property advocated by York
et. al., see e.g., \cite{Abrahams96a}.  (\ref{eq:kdot}) can be written
as
\begin{equation}
\label{eq:kdot4}
\partial_{\hat{t}} \tilde{A}^{ij} \approx (\alpha g^{-1/3})
\tilde{g}^{kl} \partial_{l} \tilde{U}^{ij}{}_{k} \approx \alpha
{g}^{kl} \partial_{l} \tilde{U}^{ij}{}_{k} ,
\end{equation}
where 
\begin{equation}
\label{eq:udefine}
\tilde{U}^{ij}{}_{k} = \frac{1}{2} ( \tilde{D}^{ij}{}_{k} - \tilde{g}^{i}_k
\tilde{D}^{il}{}_{l} - \tilde{g}^{j}_k \tilde{D}^{jl}{}_{l} +
\frac{2}{3} \tilde{g}^{ij} \tilde{g}_{km} \tilde{D}^{ml}{}_{l}) .
\end{equation}
We can take $\tilde{U}^{ij}{}_{k}$ to be our basic non-conformal
variables (note $\tilde{g} _{ij} \tilde{U}^{ij}{}_{k} =0$).  Taking the time
derivative of $\tilde{U}^{ij}{}_{k}$ and commuting time and space
derivatives leads to 
\begin{equation}
\label{eq:ddot6}
\partial_{\hat{t}} \tilde{U}^{ij}{}_{k} \approx \alpha
(\tilde{A}^{ij}{}_{,k} - \tilde{g}^i_k \tilde{A}^{jl}{}_{,l} - \tilde{g}^j_k
\tilde{A}^{il}{}_{,l} + \frac{2}{3} \tilde{g}^{ij} \tilde{g}_{km} 
\tilde{A}^{ml}{}_{,l} ) .
\end{equation}
To make the system strongly hyperbolic, 
we follow the
step leading to (\ref{eq:ddot2}) and add the combination of momentum
constraints $\alpha g^{1/3}(\tilde{g}^i_k H^j + \tilde{g}^j_k H^i) - 2
\alpha \tilde{g}^{ij} H_k /3$ to (\ref{eq:ddot6}) to arrive at
\begin{equation}
\label{eq:udot}
\partial_{\hat{t}} \tilde{U}^{ij}{}_{k} \approx \alpha
\tilde{A}^{ij}{}_{,k}  .
\end{equation}
(\ref{eq:kdot4}) and (\ref{eq:udot}) form a conformal hyperbolic
system for \{$\tilde{U}^{ij}{}_{k},\tilde{A}^{ij}$\} with only physical
characteristic speeds. The system can be symmetrized by contracting
(\ref{eq:udot}) with $g^{kl}$.

\paragraph {DISCUSSION AND CONCLUSION.}
We raise the question of what part of the variables in the Einstein
theory should be evolved in a hyperbolic fashion in numerical relativity.
We propose a re-formulation of the Einstein evolution equations that
cleanly separate the conformal degrees of freedom \{$g, g_{,i}, K$\}
and the non-conformal degrees of freedom \{$ \tilde{g}^{ij},
\tilde{D}^{ij}{}_{k}, \tilde{A}^{ij} $\} (or their linear combinations),
with the latter satisfying a first order strongly hyperbolic system.
The conformal degrees of freedom are taken to be
determined by the choice of slicings and the initial data, and are
regarded as given functions in the hyperbolic part of the evolution
equations, along with the lapse and the shift.

We find a two parameter family of non-conformal hyperbolic system
for \{$ \tilde{g}^{ij} , \tilde{D}^{ij}{}_{k}, \tilde{A}^{ij} $\}.  The
two parameters are uniquely fixed if we require the system to be
``consistently trace-free'', i.e., the time derivative of the
trace-free variables \{$ \tilde{g}^{ij} , \tilde{D}^{ij}{}_{k},
\tilde{A}^{ij} $\} remains trace-free to principal part, even in the
presence of constraint violations caused by numerical truncation
error.  We also show that certain linear combinations of the
$\tilde{D}^{ij}{}_{k}$ lead to a conformal hyperbolic system with
physical characteristic speed.

This formulation merges two recent trends in re-writing the Einstein
evolution equations for numerical relativity: first order
hyperbolicity and the separating out of the conformal degrees of
freedom.  We believe it will lead to many interesting investigations:
Given the coordinate conditions, e.g., maximal slicing and an
appropriate shift condition, can the combined elliptic hyperbolic
system be shown to be well-posed analytically \cite{Choquet95,Andersson99a}?
When posted as initial boundary value problem, what are the suitable
boundary conditions for stability in numerical evolutions?  How will
the constraints propagate under this system of conformal-hyperbolic
equations?  One particularly interesting issue that will be reported on
in a follow up paper is the stability of this formulation in numerical
evolution, and how the stability is related to the slicing conditions
($K$) one chooses.

This research is supported by the NSF grant Phy 96-00507, and NASA HPCC
Grand Challenge Award NCCS5-153.  We thank Carles Bona, Helmut
Friedrich, Alan Rendall, and Ed Seidel for discussions.


\bibliographystyle{prsty}


\end{multicols}
\end{document}